# Enhancing Bistable Vibration Energy Harvesters with Tunable Circuits: A Comparative Analysis

Adrien Morel[1], Quentin Demouron[1], Adrien Badel[1]

[1] SYMME, Université Savoie Mont-Blanc, France

E-mail: adrien.morel@univ-smb.fr

**Abstract**

In this article, we present an analysis of the effects of nonlinear circuits on bistable vibration energy harvesters. We begin by introducing an analytical model for bistable vibration energy harvesters and demonstrate that the impact of nonlinear circuits can be characterized by two factors: electrically-induced damping and electrically-induced stiffness. Subsequently, we investigate how these factors influence the power-frequency response of the harvester. Our findings reveal that electrically-induced damping significantly affects the bistable vibration energy harvester dynamics, whereas electrically-induced stiffness has minimal impact, which is a notable difference from the behavior of linear harvesters connected to nonlinear circuits. Thereafter, we conduct a comparative study of bistable energy harvesters connected to different nonlinear circuits already well documented in the literature. Our analysis demonstrates that, in most cases, the parallel synchronized switch harvesting on inductor circuit yields superior performance due to its ability to maximize electrically-induced damping. These comparative assessments and conclusions are evaluated within the framework of our proposed models and are contrasted with results obtained from linear vibration energy harvesters. The derived comparison maps presented at the end of this paper offer quantitative justification for selecting the optimal circuit for bistable energy harvesters, while deepening our comprehension of the intricate dynamics associated with nonlinear harvesters coupled with nonlinear circuits.

## 1. Introduction

In recent years, vibration energy scavenging has emerged as a promising alternative to batteries for powering sensor nodes, especially in closed confined environments where traditional energy sources like solar and thermal gradients are limited [1]. Among various electromechanical transduction mechanisms, the utilization of piezoelectric energy harvesters at the centimeter scale has gained significant attention due to their ability to convert mechanical strain into storable electrical energy with high power densities [2].

Piezoelectricity-based vibration energy harvesters (VEH) consist of two main components: i) The electromechanical resonator, which converts ambient vibration energy into strain applied to the piezoelectric material. This strain generates electrical charges due to direct piezoelectric effect. ii) The electrical circuit, responsible for extracting energy from the piezoelectric material and converting the AC voltage across its electrodes into a DC voltage suitable for charging a battery or capacitor.

Both these two aspects of VEH have been thoroughly investigated and optimized during the last decades. In order to enhance the power and harvesting bandwidth of electromechanical resonators, nonlinearities have been progressively incorporated and engineered [3]. Indeed, as proven in the seminal paper of Cottone et al. [4], nonlinear electromechanical resonators exhibit wider bandwidths of their linear counterparts, tackling one of their main drawbacks: their sensitivity to vibration frequency shifts [5]. Various types of nonlinear VEH have been studied by the community, in order to enhance the harvesting power and bandwidth, such as monostable resonators [6,7], bistable Duffing-type resonators [8,9], and more recently, tri-stable [10] and multi-stable [11] resonators.

On the other hand, significant efforts have been made to optimize the electrical circuits for VEHs with the goal of maximizing harvested power and bandwidth [12]. Initially, the standard energy harvesting (SEH) circuit, comprising a full-diode bridge followed by a DC-DC converter, was studied and optimized as a straightforward approach to adapt conventional power electronics systems for vibration energy harvesting [13, 14]. To further increase the extracted power, researchers have proposed a range of nonlinear circuits that outperform SEH. Notable examples include the synchronous electrical charge extraction (SECE) circuit [15] and the synchronized switch harvesting on inductor (SSHI) circuit [16]. These circuits have demonstrated the capability to multiply harvested power by factors ranging from 2 to 10 compared to SEH [17]. Over the years, several improved variants, such as the double synchronized switch harvesting (DSSH) [18], enhanced synchronized switch harvesting (ESSH) [19], synchronized switch harvesting on capacitors (SSHC), and synchronized multiple bias-flip (SMBF) [20] circuits, have been developed and implemented.

More recently, researchers have explored the potential of tunable circuits to enhance the bandwidth of linear VEHs by electrically adjusting the harvester dynamics [21, 22]. For instance, the frequency tuning SECE (FT-SECE) [23], short-circuit SECE (SC-SECE) [24], and phase-shifted SSHI (PS-SSHI) [25] are notable improvements of SECE and SSHI circuits that enable bandwidth enhancement of linear VEHs by a factor ranging from 2 to 6 [26]. Collectively, these advancements in electrical circuits for VEHs have played a vital role in maximizing harvested power and expanding the bandwidth of linear VEHs.

However, despite tremendous efforts in independently optimizing the electromechanical resonator and the electrical circuits, the majority of studies have focused on either analyzing a nonlinear electromechanical resonator connected to a simple resistance, which serves as an emulation of the electrical circuit, or investigating complex nonlinear circuits connected to a linear or standardized VEH. Only a few research works have delved into the interactions between nonlinear electromechanical resonators and nonlinear circuits. For example, Singh et al. [27] proposed a joint study and modeling of a bistable piezoelectric energy harvester (PEH) coupled with a synchronized switch harvesting on inductor (SSHI) circuit in 2015. Their numerical and experimental investigation demonstrated the benefits of such a combination in enhancing harvested power under broadband excitations. In 2019, Huguet et al. [28] presented an analytical and experimental study of a synchronous electrical charge extraction (SECE) circuit and a tunable SECE circuit coupled with a bistable PEH. Their work showcased improved performance compared to using a resistive load. Additionally, Wang et al. [29] proposed an analytical modeling approach in 2020 for a nonlinear monostable VEH combined with various nonlinear circuits, based on the impedance model. Although their study focused on a monostable VEH and is not directly applicable to bistable VEH, it provided valuable insights into the utility and optimization of nonlinear circuits for nonlinear VEH. These works [27-29] have shed light on the potential of integrating nonlinear electromechanical resonators and nonlinear circuits, paving the way for further exploration and understanding of their mutual influences and optimization for designing robust VEH.

In this paper, our objective is to investigate the mutual impact, influence, and optimization of nonlinear circuits in conjunction with one of the most commonly used nonlinear VEH, namely bistable VEH. The second section of the paper starts by considering the conventional electromechanical model of nonlinear VEH and demonstrates that the influence of the electrical circuit on the dynamics of nonlinear VEH can be effectively characterized by two factors: electrically-induced damping and electrically-induced stiffness. Section 3 focuses on analyzing the effects of these two factors on the dynamics of bistable VEH, with a specific emphasis on the power-frequency response. Section 4 offers insights into maximizing the harvested power from bistable VEH by strategically adjusting the electrically-induced stiffness and damping. Finally, based on the developed models and analysis, Section 5 presents a comprehensive comparison of existing nonlinear circuits for bistable VEH, providing researchers with guidelines for selecting the most suitable circuit based on the level of harvested vibration and the specific characteristics of the bistable VEH.

## 2. Electromechanical model of bistable VEH

### *2.1 Electromechanical harvester dynamics*

An example of single degree of freedom bistable VEH based on a piezoelectric transducer is shown in Figure 1. By considering the various forces applied on the inertial mass and the piezoelectric material length variation, the electromechanical dynamics of this bistable VEH can be derived [30]. The differential equations modelling the dynamics of such VEH are given by (1).

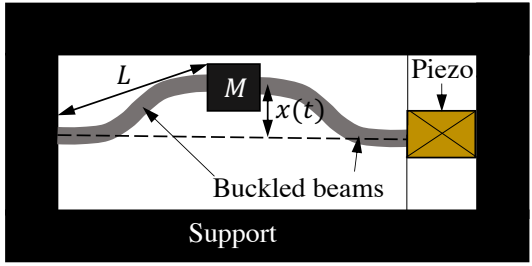

**Figure 1:** Bistable VEH based on buckled beams structure [30].

$$\begin{cases} A = \ddot{x} + \frac{\omega_0^2}{2}\left(\frac{x^2}{x_0^2} - 1\right)x + \frac{\omega_0}{Q}\dot{x} + \frac{2\alpha\, x\, v_p}{M\, L} \\ \frac{2\,\alpha\, x\, \dot{x}}{L} = C_p \dot{v}_p + i_p \end{cases} \quad (1)$$

Where $x$, $\dot{x}$ and $\ddot{x}$ represent the position, speed and acceleration of the inertial mass $M$ over time, respectively. $A$ is the acceleration of excitation. Because of the buckling of the beams, the inertial mass has two stable positions, for $x = x_0$ and $x = -x_0$. The constants $\omega_0$ and $Q$ correspond to the natural angular frequency and mechanical quality factor of the equivalent linear harvester, which is obtained for small oscillations of the mass around one of its stable positions. $L$ represents the length of one of the buckled beams, as indicated in Fig.1. As the inertial mass oscillates due to external excitation, a strain is applied on the piezoelectric material, and part of the mechanical power of the VEH is converted in electrical power. The voltage across the piezoelectric material is noted $v_p$, and the current flowing in the energy extraction interface is noted $i_p$. The parameters $\alpha$ and $C_p$ represent the force factor of the harvester and the clamped capacitance of the piezoelectric material, respectively.

## 2.2 Orbits of bistable VEH

Because of the nonlinearity of (1), multiple types of motion may exist for a given excitation [28]. In order to visualize some of these solutions, (1) has been numerically solved with the set of parameters shown in Table 1. To obtain these results, we considered that the piezoelectric material is in short-circuit configuration (i.e., $v_p = 0V$), and that the excitation is a sinusoid of amplitude $7m/s^2$ and frequency 50Hz. The obtained time-varying waveforms and the corresponding orbits in the phase space $(x, \dot{x})$ are illustrated in Fig.2.

**Table 1:** Parameters of the simulated bistable VEH

| Var. | Definition | Values | Expressions | Units |
|---|---|---|---|---|
| $M$ | Mass | 6.5 | - | g |
| $K$ | Stiffness | 349060 | - | N/m |
| $\mu_m$ | Damping | 37 | - | N/(m.s$^{-1}$) |
| $x_0$ | Stable position | 0.8 | - | mm |
| $\alpha$ | Force factor | 0.15 | - | N/V |
| $L$ | Beams length | 35 | - | mm |
| $C_p$ | Piezo. capacitance | 1 | - | $\mu$F |
| $\omega_0$ | Eq. linear natural angular frequency | 335 | $\frac{x_0}{L}\sqrt{\frac{4K}{M}}$ | rad.s$^{-1}$ |
| $Q$ | Eq. linear quality factor | 80 | $\frac{x_0}{\mu_m L}\sqrt{4KM}$ | - |
| $k_m^2$ | Expedient electromech. coupling | 0.068 | $\frac{\alpha^2}{K\, C_p}$ | - |
| $v_{oc,m}$ | Open-circuit voltage amplitude | - | $\frac{\alpha\, x_m^2}{2\, L\, C_p}$ | V |

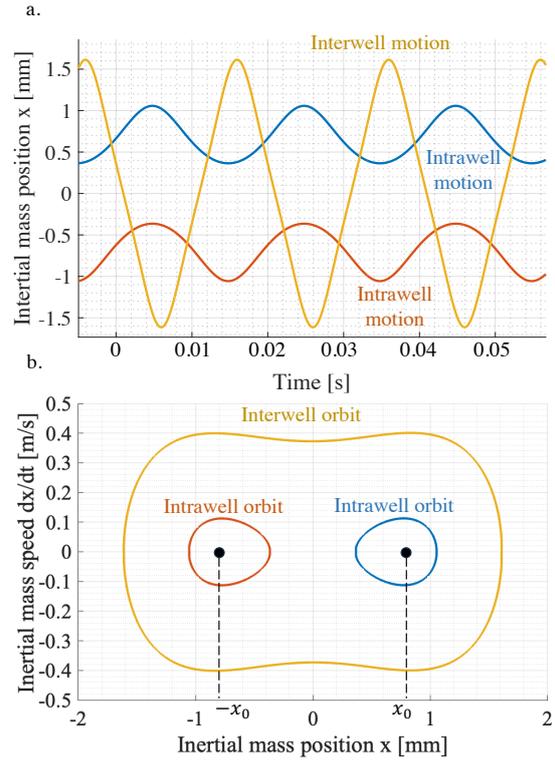

**Figure 2:** a. Time-varying waveforms and b. phase-space of bistable VEH, illustrating intrawell orbits (red and blue) and interwell orbit (yellow).

As shown in Fig.2, the largest existing orbit (in yellow) corresponds to the oscillation of the mass around the two stable positions of the VEH, with a frequency equal to the frequency of the ambient acceleration $A(t)$. This type of motion is usually called *interwell motion* in the literature. Since interwell motion allows to maximize the mechanical displacement and the harvested power, many approaches have been developed in order to jump from other less favorable orbits (such as *intrawell orbits* shown in red and blue in Fig.2) to this orbit [31, 32]. Therefore, the next sections of this paper will solely focus on the analysis and modelling of this high-energy interwell motion.

*2.3 Assumptions and first-harmonic analysis*

In order to derive compact expressions governing the behavior of the interwell motion, we will consider a few simplifying assumptions:

i) The ambient acceleration is assumed to be sinusoidal and can be written $A(t) = A_m \cos(\omega t + \psi)$, with $A_m$ the acceleration amplitude, $\psi$ its phase, and $\omega$ its angular frequency. Such assumption is reasonable in many applicative cases where the ambient vibration energy is concentrated around a single frequency [33, 34].

ii) The mechanical displacement is considered sinusoidal when the VEH operates in interwell motion. This assumption is reasonable since the harmonics in interwell motion are relatively low [35].

iii) Only the first harmonic of the piezoelectric voltage will have an impact on the dynamics of the piezoelectric energy harvester. This assumption, called *first harmonic assumption* in the literature [12], is directly related to assumption ii): the mechanical displacement is sinusoidal therefore the impact of the voltage higher harmonics on the mechanical displacement is negligible.

Combining ii) with the second equation of (1), it can be shown (with any traditional nonlinear electrical interfaces [12]) that $v_p(t)$ is a periodic signal whose fundamental frequency is $2\omega$ (due to the $\frac{2\alpha x \dot{x}}{L}$ term in the second equation of (1)). Therefore $v_p(t)$ can be expressed as a Fourier series (2).

$$v_p(t) = \sum_{n=1}^{N} [a_n \cos(2n\omega t) + b_n \sin(2n\omega t)] \quad (2)$$

Under these first harmonic assumptions ii) and iii), the mechanical displacement as well as the piezoelectric voltage can be rewritten as follows:

$$\begin{cases} x(t) \approx x_m \cos(\omega t) \\ v_p(t) \approx a_1 \cos(2\omega t) + b_1 \sin(2\omega t) \end{cases} \quad (3)$$

with $x_m$ being the displacement amplitude. $a_1$ and $b_1$ are the first coefficients of the Fourier series of $v_p(t)$, and can be determined by harmonic analysis [35].

*2.4 Electrically induced damping and stiffness*

Multiplying the first equation of (1) by $M$ and combining it with (3) yields (4):

$$\begin{cases} M\gamma = M\ddot{x} + \frac{M\omega_0^2}{2}\left(\frac{x^2}{x_0^2} - 1\right)x + \frac{M\omega_0}{Q}\dot{x} + F_p \\ F_p = \frac{2\alpha x v_p}{L} \approx \frac{\alpha a_1}{L}x - \frac{\alpha b_1}{L\omega}\dot{x} \end{cases} \quad (4)$$

with $F_p$ being the electrically induced force on the mechanical part of the bistable VEH. As shown by the second equation of (4), this force has a component in-phase with the mechanical displacement $x$, and a component in-phase with the mechanical speed $\dot{x}$. Re-arranging the terms of (4) yields (5).

$$MA = M\ddot{x} + \left(\frac{M\omega_0^2}{2}\left(\frac{x^2}{x_0^2} - 1\right) + \frac{\alpha a_1}{L}\right)x + \left(\frac{M\omega_0}{Q} - \frac{\alpha b_1}{L\omega}\right)\dot{x} \quad (5)$$

(5) proves that the electrical interface acts as an additional stiffness $K_e = \frac{\alpha a_1}{L}$ and an additional damping $\mu_e = -\frac{\alpha b_1}{L\omega}$ of electrical origin. These two terms can be rewritten as dimensionless variables, with $\beta = \frac{\mu_e}{\mu_m}$ the damping ratio, and $\nu = \frac{K_e}{M\omega_0^2}$ the stiffness ratio:

$$MA = M\ddot{x} + \left(\frac{M\omega_0^2}{2}\left(\frac{x^2}{x_0^2} - 1 + 2\nu\right)\right)x + \frac{M\omega_0}{Q}(1+\beta)\dot{x} \quad (6)$$

Therefore, the influence of any electrical interface on the dynamics of the bistable VEH is gathered in these two ratios, $\beta$ and $\nu$.

## 3. Impact of the electrical interface on the dynamics

### 3.1 Displacement amplitude $x_m$

Combining (6) with the expressions of $A(t)$ and $x(t)$, and applying the harmonic balance to the first-order terms leads to (7):

$$\begin{cases} -\omega^2 x_m - \dfrac{x_m \omega_0^2 (1-2\nu)}{2} + \dfrac{3\omega_0^2}{8x_0^2} x_m^3 = A_m \cos(\psi) \\ \dfrac{\omega_0(1+\beta)}{Q}\omega x_m = A_m \sin(\psi) \end{cases} \quad (7)$$

Squaring and adding up the two equations of (7) leads to (8).

$$\left(\dfrac{\omega^2}{\omega_0^2} + \dfrac{(1-2\nu)}{2} - \dfrac{3x_m^2}{8x_0^2}\right)^2 + \left(\dfrac{(1+\beta)\omega}{Q\,\omega_0}\right)^2 - \left(\dfrac{A_m}{x_m \omega_0^2}\right)^2 = 0 \quad (8)$$

For most harvesters in the literature, the mechanical quality factor $Q$ tends be relatively large (usually greater than 50). In the one hand, the value of the damping ratio $\beta$ usually remains limited (smaller than 1 or 2). Additionally, in interwell motion, the mass acceleration $(\omega^2 x_m)$ is usually much larger than the vibration acceleration $(A_m)$ (for instance, in [x], the mass acceleration is 30 times larger than the vibration acceleration for a vibration frequency of 50Hz). For these reasons, the second and third terms of (8) can be considered negligible. Taking this into account yields equation (9).

$$\dfrac{\omega^2}{\omega_0^2} + \dfrac{(1-2\nu)}{2} - \dfrac{3x_m^2}{8x_0^2} = 0 \quad (9)$$

Isolating $x_m$ in (9) yields the expression of the displacement amplitude given by (10).

$$x_m = \dfrac{2}{\sqrt{3}} x_0 \sqrt{\left(\dfrac{2\omega^2}{\omega_0^2} + (1-2\nu)\right)} \quad (10)$$

As proven by (10), the displacement amplitude in interwell motion grows with the vibration frequency and is proportional to the stable position of the mass, $x_0$. One might notice that the displacement-frequency response depends on the electrically induced stiffness, while the electrically-induced damping has no impact on the mechanical displacement.

### 3.2 Cut-off frequency $\omega_c$

As largely discussed in the literature, the interwell motion exists if the vibration frequency remains below a *cut-off frequency* $\omega_c$ [36]. As proven in [35] and [36], this cut-off frequency occurs when the phase-lag between the displacement $x(t)$ and the ambient acceleration $\gamma(t)$ is equal to 90°. Combining equations (7) with this condition ($\psi = 90°$) yields (11).

$$\begin{cases} -\omega_c^2 x_m - \dfrac{x_m \omega_0^2 (1-2\nu)}{2} + \dfrac{3\omega_0^2}{8x_0^2} x_m^3 = 0 \\ \dfrac{\omega_0(1+\beta)}{Q}\omega_c x_m = A_m \end{cases} \quad (11)$$

Solving the two equations of (11) leads to the expression of the cut-off frequency $\omega_c$:

$$\omega_c = \dfrac{\omega_0}{2}\sqrt{\left[-(1-2\nu) + \sqrt{(1-2\nu)^2 + \dfrac{6Q^2 A_m^2}{\omega_0^4 x_0^2 (\beta+1)^2}}\right]} \quad (12)$$

As detailed in [35], the cut-off frequency increases with a larger acceleration amplitude, or with a larger quality factor. On the other hand, a larger $x_0$ tends to decrease $\omega_c$. One might notice that the electrical interface has a strong influence on this cut-off frequency, both with the stiffness ratio $\nu$ and the damping ratio $\beta$. Hence, (12) shows that a proper electrical tuning of $\nu$ and $\beta$ could allow to adjust the interwell motion cut-off frequency and optimize the bandwidth of bistable VEH. This confirms some tendencies that have been empirically observed in the literature [28].

### 3.3 Extracted power $P_{ext}$

In order to evaluate performances of electrical circuits, a crucial parameter to model is the extracted power from the VEH, $P_{ext}$. Under the assumption of sinusoidal displacement, the expression of the extracted power corresponds to the power dissipated in the electrical damper $\mu_e$, and is given by (13).

$$P_{ext} = \dfrac{1}{T}\int_0^T \mu_e \dot{x}^2 \, dt = \dfrac{M\,\omega_0\,\beta\,\omega^2 x_m^2}{2Q} \quad (13)$$

From (10) and (13) yields the expression of the extracted power:

$$P_{ext} = \frac{2 M \omega_0 \beta \omega^2}{3Q} x_0^2 \left(\frac{2\omega^2}{\omega_0^2} + (1 - 2\nu)\right) \quad (14)$$

(14) shows that the extracted power grows with the vibration frequency (until the vibration frequency reaches the cut-off frequency of interwell motion). Note that for a given vibration frequency, if interwell motion exists, a larger damping ratio $\beta$ leads to a larger extracted power $P_{ext}$. However, a larger damping ratio also decreases the cut-off frequency of interwell motion $\omega_c$, which results in a trade-off between harvested power and bandwidth. Note that the power-frequency response of bistable VEH is impacted by both the electrically induced damping and stiffness.

In a similar fashion as linear VEH, the extracted power of bistable VEH is bounded and has a power limit $P_{lim}$ defined by (15) [35].

$$P_{lim} = \frac{M Q A_m^2}{8 \omega_0} \quad (15)$$

Whether the harvester is linear or nonlinear, this power limit can only be reached under specific conditions: the electrical damping should match the mechanical damping ($\beta = 1$) and $\omega = \omega_c$ (resp. $\omega = \omega_0$) for bistable (resp. linear) VEH.

*3.4 Discussions on the impacts of electrical interfaces*

As proven by (10), (12) and (14), the displacement amplitude, cut-off frequency, and harvested power of bistable VEH all depend on the values of the damping ratio $\beta$ and stiffness ratio $\nu$, and can therefore be electrically tuned. The impact of the electrically induced stiffness and electrically induced damping on linear VEH has already been thoroughly described in the literature [12]. However, these impacts are substantially different for bistable VEH compared to linear VEH. To study these differences, Fig.3 shows the theoretical power-frequency responses of both linear and bistable VEH, with various values of $\beta$. The linear VEH corresponds to a linearized version of the bistable VEH described by (1), and its constitutive equations are given by (16) (with the parameters given in Table1). The expressions of the displacement-frequency and power-frequency responses of linear VEH are derived in [12].

$$\begin{cases} MA = M\ddot{x} + M\omega_0^2 x + \frac{M\omega_0}{Q}\dot{x} + \alpha v_p \\ \alpha \dot{x} = C_p \dot{v}_p + i_p \end{cases} \quad (16)$$

Note that in Fig.3, 4 and 5, the extracted power is normalized with respect to the power limit $P_{lim}$ ($P_{nor} = P_{ext}/P_{lim}$), and the angular vibration frequency $\omega$ is normalized with respect to $\omega_0$ ($\Omega = \omega/\omega_0$).

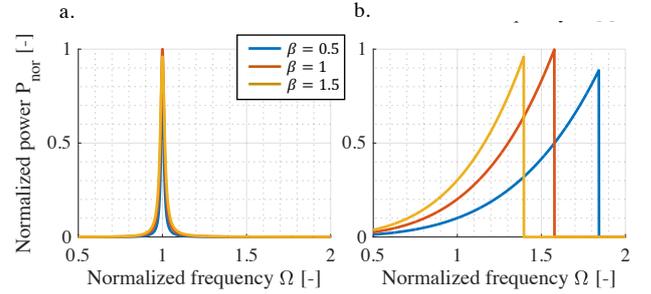

**Figure 3:** Power-frequency responses of a. linear VEH and b. bistable VEH, with three values of $\beta$, and $\nu = 0$.

Figure 3 shows illustrates that electrically induced damping does not have much impact on the power-frequency response of linear VEH. Indeed, the resonant frequency of any linear VEH does not vary with $\beta$. On the other hand, adjusting the electrically induced damping strongly impacts the power-frequency response of bistable VEH. Indeed, as proved by (12) and (14), a larger $\beta$ leads to a larger harvested power but a smaller cut-off frequency.

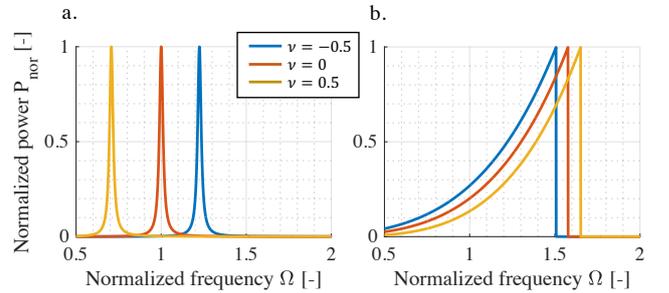

**Figure 4:** Power-frequency responses of a. linear VEH and b. bistable VEH, with three values of $\nu$, and $\beta = 1$.

Figure 4 shows the theoretical power-frequency responses of both linear and bistable VEH, with various values of $\nu$. Figure 4 illustrates that adjusting the electrically induced stiffness strongly impacts the power-frequency frequency response of linear VEH, because it tunes the VEH resonant frequency. In the other hand, a change of $\nu$ only induces a slight shift of the power-frequency response of bistable VEH. Therefore, the electrically induced stiffness has a weak impact on the power-frequency response of bistable VEH.

## 4. MPP of bistable energy harvesters

### 4.1 Maximum power-frequency responses

Figures 3 and 4 show that the power-frequency response of both bistable and linear VEH can be tuned with an electrical interface, by adjusting the electrically induced damping and electrically induced stiffness. In order to evaluate the maximum performances of a specific electrical circuit associated with a given harvester, it is possible to compute the maximum power-frequency response that can be obtained with a fine tuning of $\beta$ and $\nu$. As a matter of example, Fig.5 shows the maximum power-frequency response that can be obtained with both a linear and a bistable VEH, with $\beta \in [0, \beta_{max}]$ and $\nu \in [0, \nu_{max}]$.

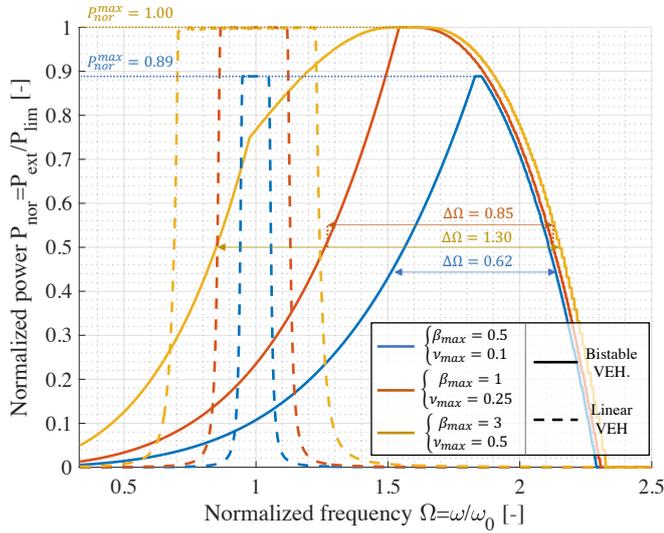

**Figure 5:** Maximum power-frequency responses of linear VEH (dashed lines) and bistable VEH (solid lines), with three combinations of $\beta_{max}$ and $\nu_{max}$.

Note that the maximum value of the damping ratio, $\beta_{max}$, and the maximum value of the stiffness ratio, $\nu_{max}$, depend on the electromechanical coupling of the harvester $k_m^2$, as well as on the choice of the electrical circuitry connected to the piezoelectric electrodes.

As shown in Fig.5, the larger $\beta_{max}$ and $\nu_{max}$, the better the performances of the harvester. Indeed, larger values of $\beta_{max}$ and $\nu_{max}$ tend to increase the maximum normalized power $P_{nor}^{max} = \max(P_{nor})$ as well as the normalized harvesting bandwidth $\Delta\Omega$, defined as the normalized frequency band where the normalized power is greater than half the maximum power.

### 4.2 Quantitative evaluation of the impact of $\beta$ and $\nu$

In order to quantitively evaluate the performances of linear and bistable VEH when connected to a given electrical circuit, Fig.6 shows the evolution of the maximum normalized power $P_{nor}^{max}$ as a function of $\beta_{max}$ and $\nu_{max}$ for linear VEH (Fig.6.a) and bistable VEH (Fig.6.b).

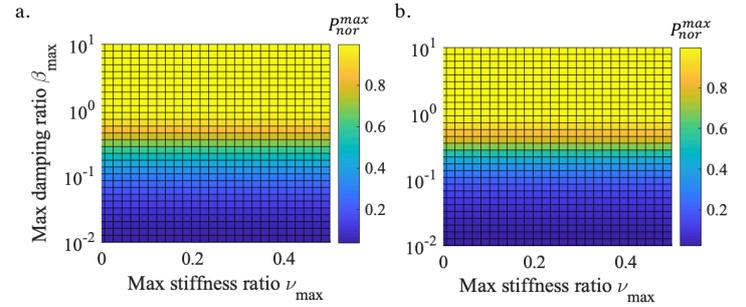

**Figure 6:** Maximum normalized power of a. linear VEH and b. bistable VEH, as a function of $\beta_{max}$ and $\nu_{max}$.

As it can be seen from Figure 6, increasing $\nu_{max}$ has little to no impact on the value of the maximum normalized power, $P_{nor}^{max}$. The only condition that must be met to reach $P_{nor}^{max} = 1$ is $\beta_{max} \geq 1$. Indeed, $P_{nor}^{max} = 1$ is obtained when the electrical damping is equal to the mechanical damping, whether the PEH is linear or bistable.

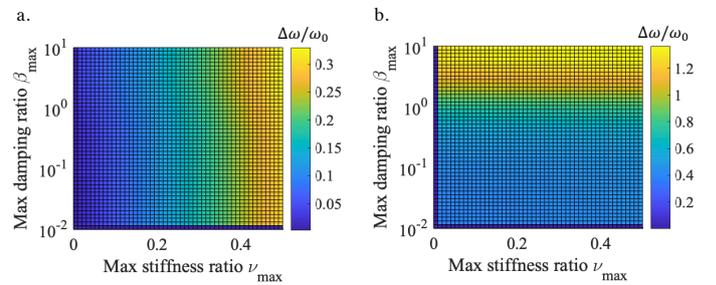

**Figure 7:** Bandwidth of a. linear VEH and b. bistable VEH, as a function of $\beta_{max}$ and $\nu_{max}$.

Figure 7 shows the evolution of the normalized harvesting bandwidth $\Delta\Omega$ as a function of $\beta_{max}$ and $\nu_{max}$ for linear VEH (Fig.7.a) and bistable VEH (Fig.7.b). For linear VEH, the bandwidth does not depend much on the value of $\beta_{max}$. Therefore, increasing the bandwidth of linear VEH requires a tuning of the stiffness ratio $\nu$.

In the case of bistable VEH, the variation of the bandwidth with $\nu_{max}$ is almost inexistant, meaning that the performances of bistable VEH exclusively depend on the electrically induced damping $\beta$. Therefore, increasing the bandwidth of bistable VEH requires a tuning of $\beta_{max}$. Note that in Fig.7, the bandwidth of bistable VEH is much larger that the bandwidth of linear VEH, which is a well-known advantage of bistable VEH [4].

Figures 6 and 7 prove that linear VEH require the maximization of two parameters ($\beta$ and $\nu$) in order to simultaneously maximize harvested power and bandwidth. Bistable VEH, in the one hand, only require the optimization of the electrically induced damping $\beta$, as their performances do not depend much on $\nu$. To complete this comparison, a quantitative criterion evaluating both the power and the bandwidth of the VEH can be defined. In this article, we choose to use $\chi_{comp}$ that has already been introduced in the literature [12]. The expression of $\chi_{comp}$ is reminded in (17).

$$\chi_{comp} = \frac{\int_0^{+\infty} \max(P_{nor}) \, d\Omega}{\int_0^{+\infty} P_{nor}{}_{\beta=1,\nu=0} \, d\Omega} \quad (17)$$

$\chi_{comp}$ corresponds to the area under the maximum power-frequency response $\left(\int_0^{+\infty} \max(P_{nor}) \, d\Omega\right)$, normalized with respect to the area under a reference power-frequency response $\left(\int_0^{+\infty} P_{nor}|_{\beta=1,\nu=0} \, d\Omega\right)$. As depicted in [12], the reference power-frequency response corresponds to a linear VEH with no adjustment of the electrical interface, i.e., a fixed value of $\beta = 1$, and a fixed value of $\nu = 0$. Figure 8 shows the evolution of $\chi_{comp}$ as a function of $\nu_{max}$ and $\beta_{max}$ for linear VEH (Fig.8.a) and bistable VEH (Fig.8.b).

In the case of linear VEH, $\chi_{comp}$ increases both with $\nu_{max}$ and $\beta_{max}$, confirming that the optimization of linear VEH require the maximization of both electrically induced stiffness and damping. In the case of bistable VEH, $\chi_{comp}$ does not vary much with $\nu_{max}$, and mainly depends on $\beta_{max}$. This confirms that the performances of bistable VEH almost solely depend on the electrically induced damping.

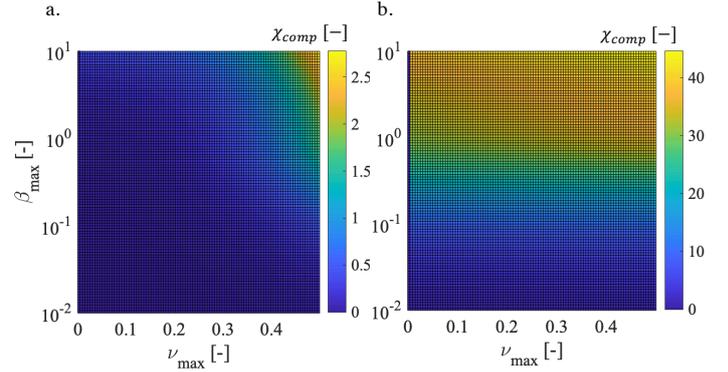

**Figure 8:** $\chi_{comp}$ of a. linear VEH and b. bistable VEH, as a function of $\beta_{max}$ and $\nu_{max}$.

The values of $\chi_{comp}$ shown in Fig.8 prove that optimizing the electrical interface with linear VEH allow to multiply the area under the power-frequency response by a factor of 3 (compared to a linear VEH with no electrical adaptation). In the case of bistable VEH, optimizing the electrical interface can lead to an area increase of 40. This last consideration proves that developing MPPT algorithms for bistable VEH is particularly relevant, while being less complex than with linear VEH (that require 2-dimensional MPPT for optimizing both $\nu$ and $\beta$).

## 5. Nonlinear circuits comparison for bistable VEH

*5.1 Dimensionless components ($\varepsilon_D, \varepsilon_K$) of the piezoelectric voltage*

As proved in Appendix B, the damping and stiffness ratio can be also expressed with dimensionless quantities:

$$\begin{cases} \beta = \frac{k_m^2 Q}{8\Omega} \left(\frac{x_m}{x_0}\right)^2 \varepsilon_D \\ \nu = \frac{k_m^2}{8} \left(\frac{x_m}{x_0}\right)^2 \varepsilon_K \end{cases} \quad (18)$$

with $\varepsilon_D = \frac{-b_1}{v_{oc,m}}$ and $\varepsilon_K = \frac{a_1}{v_{oc,m}}$ the in-phase and out-of-phase dimensionless components of the piezoelectric voltage, as defined in [12]. $a_1$ and $b_1$ correspond to the first Fourier series coefficient of the fundamental of the piezoelectric voltage whose analytical expression can be found with Fourier analysis of the time-varying piezoelectric voltage waveform. The expressions of $\varepsilon_D$ and $\varepsilon_K$ are only dependent on the choice of the electrical circuit [12]. Table 2 summarizes the expressions of $\varepsilon_D$ and $\varepsilon_K$ for a few well-known electrical circuits of the literature.

Table 2 also indicates the parameters of each electrical circuit that can be adjusted by a MPPT algorithm in order to adjust the electrically induced damping and stiffness. $r$ is the normalized load that can be adjusted by mean of a DC-DC converter (as detailed in [12]). The expression of $r$ is given by $r = 2RC_p\omega$, with $R$ being the input impedance of the DC-DC converter. Note that the expression of $r$ is two times larger in the nonlinear case compared to the linear case, because in the considered bistable VEH, the frequency of the voltage is two times larger than the vibration frequency [30].

$\gamma_{inv}$ corresponds to the voltage inversion ratio [23], $\varphi$ corresponds to the phase-shift between the energy extraction event and the maximum of the mechanical displacement [23], $\Delta\phi$ corresponds to the angular duration of a short-circuit sequence [24], and $c = \frac{C}{C+C_p}$ corresponds to the normalized capacitive load, with $c$ being a capacitance placed in parallel with the piezoelectric electrode [12]. For further information on these parameters, the reader is invited to peruse the corresponding references indicated in Table 2.

**Table 2:** Expressions of $\varepsilon_D$ and $\varepsilon_K$ for a few selected circuits from the literature. Note that the expressions of ($\varepsilon_D$, $\varepsilon_K$) remain the same for linear and nonlinear VEH, as proven in Appendix A.

| Electrical interface | Number of param. | Param. 1 | Param. 2 | $\varepsilon_D$ | $\varepsilon_K$ | Ref. |
|---|---|---|---|---|---|---|
| SECE | 0 | - | - | $4/\pi$ | 1 | [15] |
| Standard (SEH) | 1 | $r$ | - | $\dfrac{8r\Omega}{(2r\Omega + \pi)^2}$ | $\dfrac{r\Omega}{r\Omega + \pi/2}$ | [37] |
| Series SSHI | 1 | $r$ | - | $\dfrac{4}{\dfrac{\pi(1+\gamma_{min})}{(1-\gamma_{min})} + 2r\Omega}$ | 1 | [38] |
| Parallel SSHI | 1 | $r$ | - | $\dfrac{8r\Omega\left[1 + \dfrac{r\Omega(1-\gamma_{min}^2)}{2\pi}\right]}{[(1-\gamma_{min})r\Omega + \pi]^2}$ | $\dfrac{1}{1 + \dfrac{\pi}{r\Omega(1-\gamma_{min})}}$ | [37] |
| Tunable SECE | 1 | $\gamma_{inv}$ | - | $\dfrac{4}{\pi}\dfrac{1-\gamma_{inv}}{1+\gamma_{inv}}$ | 1 | [39] |
| Phase-shift SECE | 1 | $\varphi$ | - | $\dfrac{4}{\pi}\cos^2(\varphi)$ | $1 + \dfrac{2}{\pi}\sin(2\varphi)$ | [40] |
| FT-SECE | 2 | $\varphi$ | $\gamma_{inv}$ | $\dfrac{4}{\pi}\dfrac{1-\gamma_{inv}}{1+\gamma_{inv}}\cos^2(\varphi)$ | $1 + \dfrac{2}{\pi}\dfrac{1-\gamma_{inv}}{1+\gamma_{inv}}\sin(2\varphi)$ | [23] |
| SC-SECE | 2 | $\varphi$ | $\Delta\phi$ | $\dfrac{[\cos(\varphi) + \cos(\varphi+\Delta\phi)]^2}{\pi}$ $+ \sin\dfrac{(2\varphi)}{2\pi}$ $+ 2\dfrac{\cos(\varphi+\Delta\phi)\sin(\varphi)}{\pi}$ | $1 - \dfrac{\Delta\phi}{\pi} + \dfrac{\sin(2\varphi + 2\Delta\phi)}{2\pi}$ | [24] |
| PS-SSHI | 2 | $\varphi$ | $r$ | See [41] | See [41] | [41] |
| CT-SEH SC-SEH | 2 | $r$ | $c$ | $(1-c)\dfrac{8r\Omega}{(2r\Omega + \pi)^2}$ | $(1-c)\dfrac{r\Omega}{r\Omega + \pi/2}$ | [12] [42] |

## 5.2 Quantitative comparison of electrical circuits for bistable VEH

Combining the expressions of $\varepsilon_D$ and $\varepsilon_K$ in Table 2 with (12), (14), and (18) it is possible to compute the bandwidth $\Delta\Omega$, max. power $P_{nor}^{max}$, and comparison criterion. $\chi_{comp}$ for each electrical circuit, as a function of the electromechanical coupling and acceleration amplitude. Note that in this section, to accurately compare electrical circuits, we pondered the harvested power with an electrical efficiency, whose expression depends on the considered electrical circuit. The expressions of the electrical efficiencies are given and detailed in [12] and [43], and are reminded below:

$$\begin{cases} \eta_{SECE} = \eta_{PS-SECE} = \eta_{SCSECE} = \dfrac{1}{1 - \ln(-\gamma_{min})} \\ \eta_{tun.SECE} = \eta_{FTSECE} = 1 - 2\dfrac{\arccos(\gamma_{inv}) - \gamma_{inv}\sqrt{1 - \gamma_{inv}^2}}{\pi(1 - \gamma_{inv}^2)(1 - (\ln(-\gamma_{min}))^{-1})} \\ \eta_{PSSHI} = \eta_{PS-SSHI} = 2\pi\left(2\pi + \Omega r(1 - \gamma_{min}^2)\right)^{-1} \end{cases} \quad (19)$$

Based on the power expression (14), Table 2 providing electrical damping and stiffness expressions for each electrical circuits, and (19) giving the electrical efficiency of each electrical circuit, the evolution of $\chi_{comp}$ for various circuits and various VEH can be analysed. Figure 9 shows the evolution of $\chi_{comp}$ for well-known harvesting circuits, as a function of the VEH electromechanical coupling $k_m^2$, for $A_m = 10 m/s^2$.

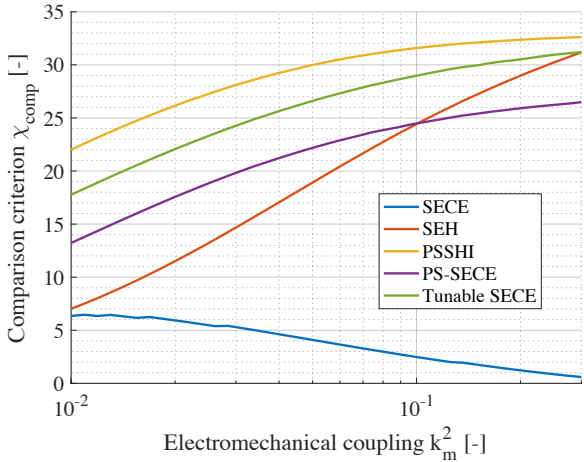

**Figure 9:** Evolution of $\chi_{comp}$ for well-known harvesting circuits, as a function of $k_m^2$, for $A_m = 10 m/s^2$.

As shown in Fig.9, the choice of the electrical circuit has a large influence of the performance of the VEH. As a matter of example, under weak electromechanical coupling ($k_m^2 = 0.01$), selecting parallel synchronized switch harvesting on inductor (PSSHI) circuit instead of standard energy harvesting (SEH) circuit allows to multiply the comparative criterion $\chi_{comp}$ by more than 3 times.

As illustrated in Fig.9, larger coupling leads to higher $\chi_{comp}$. Indeed, when the electromechanical coupling is increased, the values of $\beta_{max}$ and $\nu_{max}$ of all electrical circuits are increased (18). As proved in Fig.5-8, larger values of $\beta_{max}$ and $\nu_{max}$ lead to larger power and bandwidth, and thus larger $\chi_{comp}$. One may notice that the best electrical circuit when $A_m = 10m/s^2$ does not depend on the value of the coupling and is always the PSSHI circuit. Indeed, PSSHI circuit allows to maximize the electrically-induced damping $\beta_{max}$ but does not have much impact on the electrically-induced stiffness $\nu_{max}$ [12]. While this explains why PSSHI circuit is not always the best with linear VEH (because it does not allow for frequency tuning), the fact it maximizes $\beta_{max}$ makes it the best choice for nonlinear VEH.

Because of the nonlinearity of the energy harvester, these conclusions might vary with the ambient level of acceleration. In order to take this factor in consideration, Fig.10 and Fig.11 show the best electrical circuit for bistable VEH in the plane $(k_m^2, A_m)$, for single parameter circuits (Fig.10) and multiple parameters circuits (Fig.11).

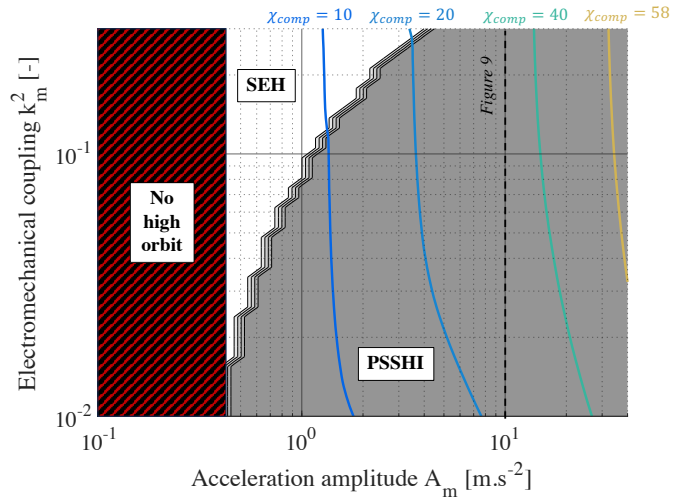

**Figure 10:** Single parameter circuits maximizing $\chi_{comp}$ depending on the VEH electromechanical coupling $k_m^2$ and acceleration amplitude $A_m$.

Figure 10 demonstrates that the PSSHI circuit remains the most optimal choice for large acceleration amplitudes and weak to moderate coupling of VEH. However, this conclusion is altered in the case of low acceleration amplitudes and strong electromechanical coupling (e.g., when $k_m^2 > 0.1$ and $A_m < 1\text{m/s}^2$). In such instances, the increased damping offered by the PSSHI circuit does not yield significant advantages compared to other circuits, such as the standard energy harvesting (SEH) circuit. This is because the VEH system is already well coupled and does not require additional electrical damping for maximizing the bandwidth and power. In the other hand, PSSHI circuit brings additional losses and complexity compared to SEH circuit, making suboptimal in this case. As illustrated in Fig.10, if the acceleration is below a threshold level (0.4m/s$^2$), the VEH interwell motion stops existing, meaning that the proposed analysis does not hold.

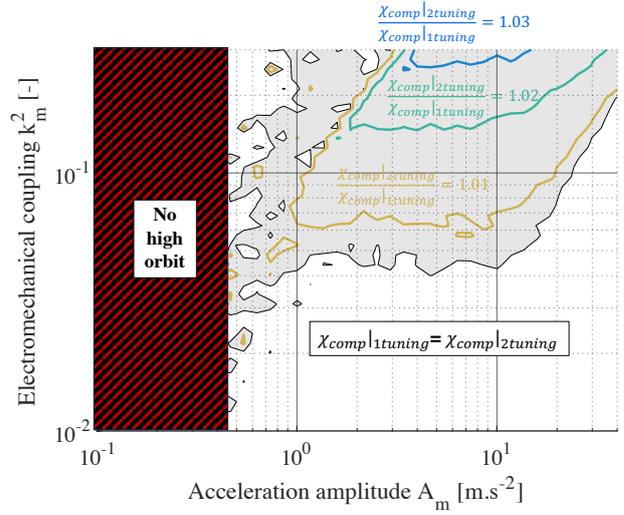

**Figure 12:** Interest of multi-tuning circuits compared to single-tuning circuits, in the $(k_m^2, A_m)$ map.

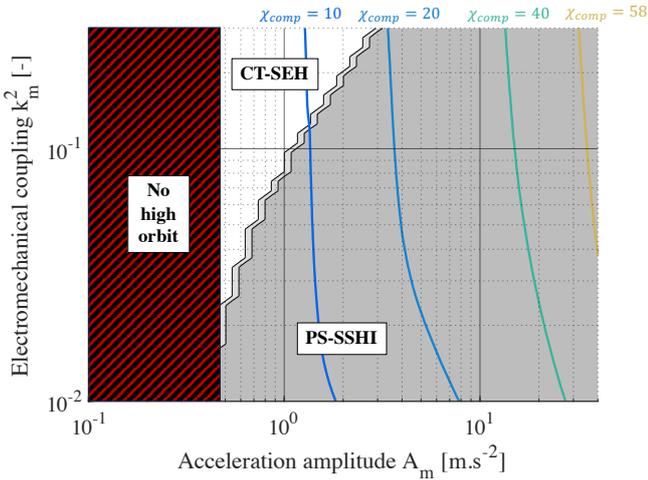

**Figure 11:** Multiple parameters circuits maximizing $\chi_{comp}$ depending on the VEH electromechanical coupling $k_m^2$ and acceleration amplitude $A_m$.

Figure 11 shows the same map considering the implementation of multiple tuning electrical circuits, such as PS-SSHI, CT-SEH, SC-SECE, FT-SECE. These circuits incorporate multiple tuning capabilities, enabling independent control of electrical damping $\beta$ and electrical stiffness $\nu$. Previous studies have demonstrated the relevance of multiple tuning circuits, especially in conjunction with linear VEH [12]. Interestingly, in the case of bistable VEH, the results closely resemble those obtained in Fig. 11. Specifically, PSSHI-type circuits (e.g., PS-SSHI) are the preferred choice, except under conditions of strong coupling and weak acceleration. In this specific scenario, SEH-type circuits (e.g., CT-SEH), with lower complexity and higher electrical efficiency, yield superior performance.

Figure 12 provides a comprehensive comparison between single tuning circuits (PSSHI, SEH, PS-SECE, etc.) and multiple tuning circuits (PS-SSHI, CT-SEH, FT-SECE, SC-SECE) when applied to bistable VEH. The white area depicted in Figure 12 represents the range of coupling and acceleration amplitudes where employing a multiple tuning circuit offers no advantage over a single tuning circuit ($\chi_{comp}|_{1tuning} = \chi_{comp}|_{2tuning}$). The gold, green, and blue lines demonstrate that the benefits provided by a multiple tuning circuit remain limited in all cases, resulting in a maximum improvement of only 3%. Indeed, in the case of bistable VEH, adjusting the electrical stiffness offers minimal advantages in terms of power and bandwidth (see Fig. 8). Consequently, a single tuning strategy (used to adjust the electrical damping) is sufficient to fully exploit the potential of bistable VEH. This conclusion stands in stark contrast to linear VEH, where multiple tuning circuits significantly enhance performance and allow for the optimization of their performances [12].

*5.3 Comparison with electrical circuits for linear VEH*

While the previous section focused on comparing nonlinear circuits for bistable VEH, it did not directly compare the performance of linear and nonlinear VEH when paired with optimized circuits. The optimization of nonlinear circuits with linear VEH has been previously investigated in [12], demonstrating the benefits of utilizing multiple tuning circuits like FT-SECE and SC-SECE. However, in the case of nonlinear VEH, the previous section showed that single tuning circuits such as PSSHI and SEH are generally sufficient to achieve optimized performance. Figure 13 presents a comparison of the performance between nonlinear VEH and linear VEH when combined with optimal circuits in the $(A_m, k_m^2)$ plane.

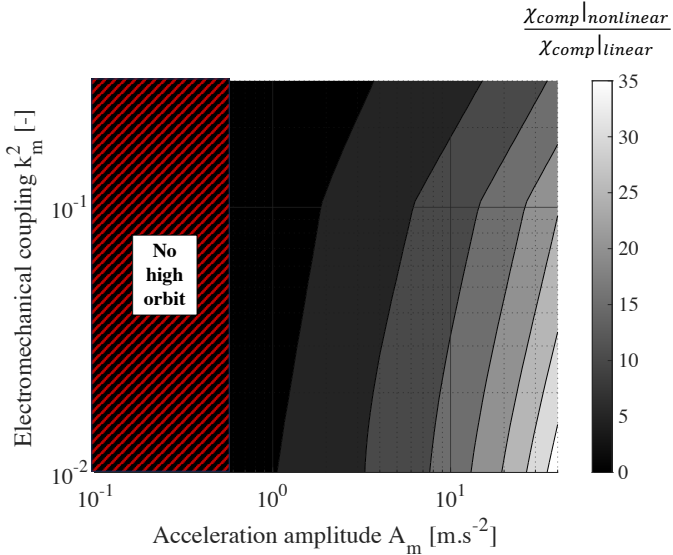

**Figure 13:** Comparison of performances of linear and nonlinear VEH associated with optimal circuits.

In Fig. 13, each combination of acceleration amplitude and electromechanical coupling is represented by the ratio of the comparison criterion $\chi_{comp}$ of nonlinear VEH to linear VEH when connected to optimal circuitry. The results demonstrate that when the acceleration amplitude is small (<1m/s$^2$) and the electromechanical coupling is relatively large, linear VEH can approach the performance level of nonlinear VEH. However, for larger acceleration amplitudes and weaker electromechanical coupling, the performance of nonlinear VEH significantly surpasses that of linear VEH (up to 35 times higher when $A_m = 40 m/s^2$ and $k_m^2 = 0.01$). This illustrates that the harvesting bandwidth of nonlinear VEH is directly influenced by vibration amplitude, while the performance of linear VEH is strongly dependent on the coupling of the VEH. Indeed, linear VEH require multi-tuning circuits exploiting the backward coupling of VEH to compensate for their inherently lower bandwidth.

## 5. Conclusion

This study has provided a comprehensive analysis of the effects of nonlinear circuits on bistable VEH. By introducing an analytical model and investigating the influences of electrically-induced damping and stiffness, valuable insights into the dynamics of these energy harvesters in the presence of nonlinear circuits have been unraveled.

Electrically-induced damping plays a significant role in shaping the behavior of bistable vibration energy harvesters, while electrically-induced stiffness has minimal impact, distinguishing them from linear harvesters connected to nonlinear circuits. This understanding opens up new avenues for optimizing the design and operation of such harvesters, ultimately enhancing their performance in energy harvesting applications.

The comparative study of various nonlinear circuits revealed that the parallel synchronized switch harvesting on inductor circuit consistently outperforms others in maximizing electrically-induced damping. This circuit has proven to be a promising choice for bistable VEH, further substantiating its potential for achieving efficient energy conversion. The comparison maps derived from this analysis provide a quantitative basis for selecting the optimal circuit for bistable VEH, empowering scientists to make informed decisions in their design choices. This study not only advances the understanding of nonlinear VEH, but also paves the way for the design of robust systems capable of harnessing relatively large power across wide frequency bands.

## References

[1] R. Vullers, R. vanSchaijk, I. Doms, C. vanHoof, and R. Mertens, "Micropower energy harvesting," *Solid-State Electron.*, vol. 53, no. 7, pp. 684–693, 2009.
[2] S. Roundy, et al. "Improving power output for vibration-based energy scavengers," *IEEE Pervasive Computing*, vol. 4, no. 1, pp. 28–36, Jan. 2005.
[3] Y. Jia, "Review of nonlinear vibration energy harvesting: Duffing, bistability, parameteric, stochastic and others," *Journal of Intelligent Material Systems and Structures*, vol. 31, no. 7, 2020.
[4] F. Cottone, H. Vocca, and L. Gammaitoni, "Nonlinear energy harvesting," *Physical Review Letters*, vol. 102, 080601, 2009.
[5] W. Liu, et al., "A new figure of merit for wideband vibration energy harvesters," *Smart Materials and Structures*, vol. 24, no. 12, 2015.
[6] N. Ngoc Linh, et al., "Efficiency of mono-stable piezoelectric Duffing energy harvester in the secondary resonances by averaging method, Part 2: Super-harmonic resonance," *International Journal of Non-linear mechanics*, vol. 137, 103817, 2021.
[7] K. Fan, et al., "A monostable piezoelectric energy harvester for broadband low-level excitations," *Applied Physics Letters*, vol. 112,

no. 12, 123901, 2018.
[8] S. Pellegrini, et al., "Bistable vibration energy harvesters: A review," *Journal of Intelligent Material Systems and Structures*, vol. 24, no. 11, pp. 1303-1312, 2013.
[9] D. Tu et al., "A bistable energy harvester with low base-acceleration and high root mean square output for train bogies: theoretical modeling and experimental validation," *Smart Materials and Structures*, vol. 32, no. 3, 035009, 2023.
[10] C. Liu et al., "Large stroke tri-stable vibration energy harvester: Modelling and experimental validation," *Mechanical Systems and Signal Processing*, vol. 168, 108699, 2022.
[11] S. Zhou, M. Lallart, A. Erturk, "Multistable vibration energy harvesters: Principle, progress, and perspectives," *Journal of Sound and Vibration*, vol. 428, 116886, 2022.
[12] A. Morel, et al., "A comparative study of electrical interfaces for tunable piezoelectric vibration energy harvesting," *Smart Materials and Structures*, vol. 31, no. 4, 045016, 2022.
[13] G. K. Ottman, et al., "Adaptive piezoelectric energy harvesting circuit for wireless remote power supply," *IEEE Transactions on Power Electronics*, vol. 17, no. 5, pp. 669-676, 2002.
[14] G. K. Ottman, H.F. Hofmann, and G.A. Lesieutre, "Optimized piezoelectric energy harvesting circuit using step-down converter in discontinuous conduction mode," *IEEE Transactions on Power Electronics*, vol. 18, no. 2, pp. 696-703, 2003.
[15] E. Lefeuvre et al., "Piezoelectric energy harvesting device optimization by synchronous electric charge extraction," *Journal of Intelligent Material Systems and Structures*, vol. 16, no. 10, pp. 865-876, 2005.
[16] D. Guyomar et al., "Toward energy harvesting using active materials and conversion improvement by nonlinear processing," *IEEE Transactions on ultrasonics, ferroelectrics and frequency control*, vol. 52, no. 4, pp. 584-595, 2005.
[17] E. Lefeuvre et al., "A comparison between several vibration-powered piezoelectric generators for standalone systems," *Sensors and Actuators A: Physical*, vol. 126, no. 2, pp. 405-416, 2006.
[18] M. Lallart, et al., "Double synchronized switch harvesting (DSSH): a new energy harvesting scheme for efficient energy extraction," *IEEE Transactions on Ultrasonics, Ferroelectrics, and Frequency Control*, vol. 55, no. 10, pp. 2119-2130, 2008.
[19] H. Shen, et al., "Enhanced synchronized switch harvesting: A new energy harvesting scheme for efficient energy extraction," *Smart Materials and Structures*, vol. 19, no. 11, 115017, 2010.
[20] J. Liang, Y. Zhao, K. Zhao, "Synchronized triple bias-flip interface circuit for piezoelectric energy harvesting enhancement," *IEEE Transactions on Power Electronics*, vol. 34, no. 1, pp. 275-286, 2018.
[21] Y. Liao and H.A. Sodano, "Optimal parameters and power characteristics of piezoelectric energy harvesters with an RC circuit," *Smart Materials and Structures*, vol. 18, no. 4, 045011, 2009.
[22] B. Zhao and J. Liang, "Circuit solutions towards broadband piezoelectric energy harvesting: an impedance analysis," *arXiv preprint*, 2022.
[23] A. Brenes, et al., "Large-bandwidth piezoelectric energy harvesting with frequency-tuning synchronized electric charge extraction," *Sensors and Actuators A: Physical*, vol. 302, 111759, 2020.
[24] A. Morel, et al., "Frequency tuning of piezoelectric energy harvesters thanks to a short-circuit synchronous electric charge extraction," *Smart Materials and Structures*, vol. 28, 025009, 2018.
[25] B. Zhao, J. Liang, and K. Zhao, "Phase-variable control of parallel synchronized triple bias-flips interface circuit toward broadband piezoelectric energy harvesting," *2018 IEEE ISCAS*, 2018.
[26] A. Morel et al., "32.2 Self-tunable phase-shifted SECE piezoelectric energy-harvesting IC with a 30nW MPPT achieving 446% energy-bandwidth improvement and 94% efficiency," *IEEE ISSCC 2020*, pp. 488-490, 2020.
[27] K. A. Singh, R. Kumar, and R. J. Weber, "A Broadband Bistable Piezoelectric Energy Harvester With Nonlinear High-Power Extraction," *IEEE Transactions on Power Electronics*, vol. 30, no. 12, pp. 6763-6774, 2015.
[28] T. Huguet, M. Lallart, and A. Badel, "Bistable vibration energy harvester and SECE circuit: exploring their mutual influence," *Nonlinear dynamics*, vol. 97, pp. 485-501, 2019.
[29] J. Wang, B. Zhao, W.-H. Liao, and J. Liang, "New insight into piezoelectric energy harvesting with mechanical and electrical nonlinearities," *Smart Materials and Structures*, vol. 29, no. 4, p. 04LT01, 2020.
[30] W. Liu, F. Formosa, A. Badel, "Optimization study of a piezoelectric bistable generator with doubled voltage frequency using harmonic balance method", *Journal of Intelligent Material Systems and Structures*, vol. 28, no. 5, p. 671-686, 2017.
[31] C. Saint-Martin et al., "Optimized and Robust Orbit Jump for Nonlinear Vibration Energy Harvesting," *ArXiv preprint*, 2023.
[32] Y. Huang, Z. Zhao, and W. Liu, "Systematic adjustment strategy of a nonlinear beam generator for high-energy orbit," *Mechanical Systems and Signal Processing*, vol. 166, 108444, 2022.
[33] V. G. Cleante, M.J. Brennan, G. Gatti, D.J. Thompson, "On the target frequency for harvesting energy from track vibrations due to passing trains", *Mechanical Systems and Signal Processing*, vol. 114, p. 212-223, 2019.
[34] S. Hanly, "Vibration analysis: FFT, PSD, and Spectrogram Basics", 2016.
[35] A. Morel, et al. "Simple analytical models and analysis of bistable vibration energy harvesters," *Smart Materials and Structures*, vol. 31, no. 10, 105016, 2022.
[36] R. L. Harne, M. Thota, and K. W. Wang, "Concise and high-fidelity predictive criteria for maximizing performance and robustness of bistable energy harvesters," *Appl. Phys. Lett.*, vol. 102, no 5, 053903, 2013.
[37] Y. C. Shu, I. C. Lien, and W. J. Wu, "An improved analysis of the SSHI interface in piezoelectric energy harvesting," *Smart Materials and Structures*, vol. 16, no. 6, 2253, 2007.
[38] I. C. Lien et al., "Revisit of series-SSHI with comparisons to other interfacing circuits in piezoelectric energy harvesting," *Smart Materials and Structures*, vol. 19, no. 12, 125009, 2010.
[39] A. Richter, et al., "Tunable interface for piezoelectric energy harvesting," *2014 IEEE 11th International Multi-Conference on Systems, Signals & Devices (SSD14)*, pp. 1-5, 2014.
[40] E. Lefeuvre, et al., "Power and frequency bandwidth improvement of piezoelectric energy harvesting devices using phase-shifted synchronous electric charge extraction interface circuit," *Journal of Intelligent Material Systems and Structures*, vol. 28, no. 20, 2017.
[41] P. H. Hsieh, C.-H. Chen, and H.-C. Chen, "Improving the scavenged power of nonlinear piezoelectric energy harvesting interface at off-resonance by introducing switching delay," *IEEE Transactions on Power Electronics*, vol. 30, pp. 3142-3155, 2015.
[42] A. Morel, et al., "Active AC/DC control for wideband piezoelectric energy harvesting," *Journal of Physics: Conference Series 773*, vol. 1, 012059, 2016.
[43] A. Morel, et al., "Electrical efficiency of SECE-based interfaces for piezoelectric vibration energy harvesting," *Smart Materials and Structures*, vol. 31, no. 1, 01LT01, 2021.

# Appendix A – Relations between $(\varepsilon_D, \varepsilon_K)$ for linear and nonlinear vibration energy harvesters

This section proposes to derive the relation between the dimensionless in-phase and out-of-phase components of the piezoelectric voltage, $\varepsilon_D$ and $\varepsilon_K$, for linear and nonlinear VEH. As the expressions of $(\varepsilon_D, \varepsilon_K)$ are already known and derived for linear VEH [12], we will use these relations to derive the expressions of all $(\varepsilon_D, \varepsilon_K)$ for nonlinear VEH.

The piezoelectric voltage during a period of vibration can be splitted in constant-voltage terms and constant-charge terms:

$$\forall \theta \in [0, 2\pi],\ v_p(\theta) = \sum_i v_{p_{CC,i}}(\theta) + \sum_j v_{p_{CV,j}}(\theta) \quad (20)$$

With $v_{p_{CC,i}}$ the piezoelectric voltage during the $i^{th}$ phase of constant charge, and $v_{p_{CV,j}}$ the piezoelectric voltage during the $j^{th}$ phase of constant voltage. This decomposition is valid for all traditional electrical circuits in the literature, such as SECE, SSHI, SC-SECE, SEH, …

### I) Linear VEH case

In the case of a linear VEH, the current due to piezoelectric effect $\alpha \dot{x}$ either flows in the electrical circuit ($i_p$) or in the piezoelectric material capacitor ($C_p \dot{v}_p$) [12]. Therefore, the constitutive electrical equation of linear VEH is as follow:

$$\alpha \dot{x} = C_p \dot{v}_p + i_p \quad (21)$$

The expression of the voltage $v_{p_{CC,i}}^{lin}$ during the $i^{th}$ constant charge phase, happening between angles $\theta_i$ and $\theta_i + \Delta \theta_i$, can be found from (21), with $i_p = 0$.

$$\begin{cases} \forall \theta \in [\theta_i, \theta_i + \Delta \theta_i], v_{p_{CC,i}}^{lin} = \alpha_{V,i} v_{oc,m}^{lin} + \int_{\theta_i}^{\theta_i + \Delta \theta_i} \frac{\alpha \dot{x}}{C_p} d\theta \\ \forall \theta \in [0, \theta_i[ \cup ]\theta_i + \Delta \theta_i, 2\pi],\ v_{p_{CC,i}}^{lin} = 0 \end{cases} \quad (22)$$

With $v_{oc,m}^{lin} = \frac{\alpha x_m}{C_p}$ the amplitude of the open-circuit piezoelectric voltage, and $\alpha_{V,i} \in \mathbb{R}$ a coefficient modelling the initial voltage at the beginning of the $i^{th}$ constant charge phase. Solving (22) leads to (23).

$$\begin{cases} \forall \theta \in [\theta_i, \theta_i + \Delta \theta_i], v_{p_{CC,i}}^{lin} = v_{oc,m}^{lin}\left(\alpha_{V,i} + [\cos(\theta)]_{\theta_i}^{\theta_i + \Delta \theta_i}\right) \\ \forall \theta \in [0, \theta_i[ \cup ]\theta_i + \Delta \theta_i, 2\pi],\ v_{p_{CC,i}}^{lin} = 0 \end{cases} \quad (23)$$

The expression of the voltage $v_{p_{CV,j}}^{lin}$ during the $j^{th}$ constant voltage phase, happening between $\theta_j$ and $\theta_j + \Delta \theta_j$, can be found from (21), with $v_p = v_{dc}^{lin}$, a constant voltage. Due to charge conservation, this constant voltage is equal to the average current flowing in the circuit output resistance, as explained in detailed papers on SSHI or SEH [37, 38]. Therefore, it yields the following equation:

$$\begin{cases} \forall \theta \in [\theta_j, \theta_j + \Delta \theta_j], v_{p_{CV,j}}^{lin} = v_{dc} = \frac{R}{2\pi}\int_{\theta_i}^{\theta_i + \Delta \theta_i} i_p d\theta \\ \forall \theta \in [0, \theta_j[ \cup ]\theta_j + \Delta \theta_j, 2\pi],\ v_p|_{CV,j}^{lin} = 0 \end{cases} \quad (24)$$

Combining (21) with (24):

$$\begin{cases} \forall \theta \in [\theta_j, \theta_j + \Delta \theta_j], v_{p_{CV,j}}^{lin} = v_{dc}^{lin} = v_{oc,m}^{lin} \frac{r\Omega}{2\pi}[\cos(\theta)]_{\theta_i}^{\theta_i + \Delta \theta_i} \\ \forall \theta \in [0, \theta_j[ \cup ]\theta_j + \Delta \theta_j, 2\pi],\ v_{p_{CV,j}}^{lin} = 0 \end{cases} \quad (25)$$

with $r = RC_p\omega_0$ and $\Omega = \omega/\omega_0$ the dimensionless resistive load and frequency, in the linear VEH case.

### II) Nonlinear VEH case

In the nonlinear case, the expression of the voltage during the $i^{th}$ constant charge phase, happening between $\frac{\theta_i}{2}$ and $\frac{\theta_i}{2} + \frac{\Delta \theta_i}{2}$, can be found from the second equation of (1), with $i_p = 0$. Note that the angles are chosen two times smaller than in the linear case because the fundamental frequency of the piezoelectric voltage is two times greater than the ambient vibration frequency, due to the nonlinearities [12].

$$\begin{cases} \forall \theta \in \left[\frac{\theta_i}{2}, \frac{\theta_i}{2} + \frac{\Delta \theta_i}{2}\right], v_{p_{CC,i}}^{nlin} = \alpha_{V,i} v_{oc,m}^{nlin} + \int_{\frac{\theta_i}{2}}^{\frac{\theta_i}{2}+\frac{\Delta \theta_i}{2}} \frac{2\alpha x \dot{x}}{C_p L} d\theta \\ \forall \theta \in \left[0, \frac{\theta_i}{2}\right[ \cup \left]\frac{\theta_i}{2} + \frac{\Delta \theta_i}{2}, \pi\right],\ v_{p_{CC,i}}^{nlin} = 0 \end{cases} \quad (26)$$

With $v_{oc,m}^{nlin} = \frac{\alpha x_m^2}{2LC_p}$ the amplitude of the open-circuit piezoelectric voltage. Solving (26) leads to (27).

$$\begin{cases} \forall \theta \in \left[\frac{\theta_i}{2}, \frac{\theta_i}{2} + \frac{\Delta \theta_i}{2}\right], v_{p_{CC,i}}^{nlin} = v_{oc,m}^{nlin}\left(\alpha_{V,i} + [\cos(2\theta)]_{\frac{\theta_i}{2}}^{\frac{\theta_i}{2}+\frac{\Delta \theta_i}{2}}\right) \\ \forall t \in \left[0, \frac{\theta_i}{2}\right[ \cup \left]\frac{\theta_i}{2} + \frac{\Delta \theta_i}{2}, \pi\right],\ v_{p_{CC,i}}^{nlin} = 0 \end{cases} \quad (27)$$

Similarly as in the linear case, the expression of the voltage $v_{p_{CV,j}}^{nlin}$ during the $j^{th}$ constant voltage phase, happening between $\frac{\theta_j}{2}$ and $\frac{\theta_j}{2} + \frac{\Delta \theta_j}{2}$, can be found from the second equation of (1), with $v_p = v_{dc}^{nlin}$, a constant voltage. Therefore, it yields the following equation:

$$\begin{cases} \forall \theta \in \left[\frac{\theta_j}{2}, \frac{\theta_j}{2} + \frac{\Delta \theta_j}{2}\right], v_{p_{CV,j}}^{nlin} = v_{dc}^{nlin} = \frac{R}{\pi}\int_{\frac{\theta_i}{2}}^{\frac{\theta_i}{2}+\frac{\Delta \theta_i}{2}} i_p d\theta \\ \forall \theta \in \left[0, \frac{\theta_j}{2}\right[ \cup \left]\frac{\theta_j}{2} + \frac{\Delta \theta_j}{2}, \pi\right],\ v_{p_{CV,j}}^{nlin} = 0 \end{cases} \quad (28)$$

Combining (1) with (28) leads to (29).

$$\begin{cases} \forall \theta \in \left[\frac{\theta_j}{2}, \frac{\theta_j}{2} + \frac{\Delta\theta_j}{2}\right], v_{p_{CV,j}}^{nlin} = v_{oc,m}^{nlin} \frac{r\Omega}{2\pi} [\cos(2\theta)]_{\frac{\theta_j}{2}}^{\frac{\theta_j}{2} + \frac{\Delta\theta_j}{2}} \\ \forall \theta \in \left[0, \frac{\theta_j}{2}\right] \cup \left[\frac{\theta_j}{2} + \frac{\Delta\theta_j}{2}, \pi\right], v_{p_{CV,j}}^{nlin} = 0 \end{cases} \quad (29)$$

with $r = 2RC_p\omega_0$ and $\Omega = \omega/\omega_0$ the dimensionless resistive load and frequency, in the nonlinear VEH case.

### III) Relations between linear and nonlinear VEH cases

As proven by equations (23), (25), (27) and (29), there exists relations between linear and nonlinear expressions of the voltages, both in constant voltage and constant charge operations:

$$\begin{cases} v_{p_{CC,i}}^{nlin} = \frac{x_m}{2L} v_{p_{CC,i}}^{lin} \\ v_{p_{CV,j}}^{nlin} = \frac{x_m}{2L} v_{p_{CV,j}}^{lin} \end{cases} \quad (30)$$

Therefore, combining (30) with the general expression of the piezoelectric voltage (20) yields:

$$v_p^{nlin} = \sum_i v_{p_{CC,i}}^{nlin} + \sum_j v_{p_{CV,j}}^{nlin}$$

$$v_p^{nlin} = \frac{x_m}{2L} \sum_i v_{p_{CC,i}}^{lin} + \frac{x_m}{2L} \sum_j v_{p_{CV,j}}^{lin} \quad (31)$$

$$v_p^{nlin} = \frac{x_m}{2L} \left( \sum_i v_{p_{CC,i}}^{lin} + \sum_j v_{p_{CV,j}}^{lin} \right)$$

$$v_p^{nlin} = \frac{x_m}{2L} v_p^{lin}$$

This relation between piezoelectric voltages is therefore simply given by the linear expression (31). Note that the coefficient $\frac{x_m}{2L}$ is also found when comparing the amplitudes of the open-circuit piezoelectric voltages for linear ($\frac{\alpha x_m}{C_p}$) and nonlinear ($\frac{\alpha x_m^2}{2LC_p}$) VEH. From (31), it is easy to prove that the Fourier coefficients of the voltages follow the same relations:

$$\forall n \in \mathbb{N}^* \begin{cases} a_n^{nlin} = \frac{x_m}{2L} a_n^{lin} \\ b_n^{nlin} = \frac{x_m}{2L} b_n^{lin} \end{cases} \quad (32)$$

Considering the dimensionless piezoelectric voltage in-phase and out-of-phase components, $\varepsilon_D$ and $\varepsilon_K$ (defined in [12]), the following relations can be proven from (31) and (32):

$$\begin{cases} \varepsilon_D^{nlin} \stackrel{\text{def}}{=} \frac{-b_1^{nlin}}{v_{oc,m}^{nlin}} = \frac{-b_1^{lin}}{v_{oc,m}^{lin}} \stackrel{\text{def}}{=} \varepsilon_D^{lin} \\ \varepsilon_K^{nlin} \stackrel{\text{def}}{=} \frac{a_1^{nlin}}{v_{oc,m}^{nlin}} = \frac{a_1^{lin}}{v_{oc,m}^{lin}} \stackrel{\text{def}}{=} \varepsilon_K^{lin} \end{cases} \quad (33)$$

This proves that $\varepsilon_D$ and $\varepsilon_K$ share the same expressions when analysing the influences of circuits on linear and nonlinear VEH. Therefore, the expressions shown in table 2 have been directly derived from the literature, where the circuits have been associated with linear VEH, but are used (in the present paper) for analysing the same circuits with nonlinear VEH.

### Appendix B – Relations between $(\varepsilon_D, \varepsilon_K)$ and $(\beta, \nu)$ for nonlinear VEH

As proven by (5), the expressions of the electrically-induced stiffness $K_e$ and damping $\mu_e$ are obtained as follows:

$$\begin{cases} K_e = \frac{\alpha}{L} a_1 \\ \mu_e = -\frac{\alpha}{L\omega} b_1 \end{cases} \quad (34)$$

Therefore, the expressions of the damping ratio $\beta$ and stiffness ratio $\nu$ are obtained as follows:

$$\begin{cases} \nu \stackrel{\text{def}}{=} \frac{K_e}{M\omega^2} = \frac{\alpha}{LM\omega^2} a_1 \\ \beta \stackrel{\text{def}}{=} \frac{\mu_e}{\mu_m} = -\frac{\alpha}{L\mu_m\omega} b_1 \end{cases} \quad (35)$$

Combining (35) with (33) leads to the relations between $(\varepsilon_D, \varepsilon_K)$ and $(\beta, \nu)$:

$$\begin{cases} \nu = \frac{\alpha}{LM\omega^2} v_{oc,m}^{lin} \varepsilon_K \\ \beta = \frac{\alpha}{L\mu_m\omega} v_{oc,m}^{lin} \varepsilon_D \end{cases} \quad (36)$$

Replacing $v_{oc,m}^{lin}$ by its expression and reformulating the obtained expressions with dimensionless terms (Table 1), relations (37) can be obtained:

$$\begin{cases} \nu = \frac{k_m^2}{8} \frac{x_m^2}{x_0^2} \varepsilon_K \\ \beta = \frac{k_m^2 Q}{8\Omega} \frac{x_m^2}{x_0^2} \varepsilon_D \end{cases} \quad (37)$$